\documentclass[conference]{IEEEtran}
\IEEEoverridecommandlockouts

\usepackage{xcolor}
\usepackage{graphicx}
\usepackage[linesnumbered,ruled]{algorithm2e}
\usepackage{adjustbox}
\usepackage{multirow}
\usepackage{cite}
\usepackage{amsmath,amssymb,amsfonts,amsthm}
\usepackage{textcomp}
\usepackage{cleveref}
\usepackage{caption}
\usepackage{subcaption}
\usepackage{url}

\crefname{thm}{Theorem}{Theorems}
\crefname{lem}{Lemma}{Lemmas}
\crefname{defn}{Definition}{Definitions}
\crefname{figure}{Fig.}{Figs.}
\crefname{table}{Table}{Tables}
\crefname{algorithm}{Algorithm}{Algorithms}
\crefname{equation}{Eq.}{Eqs.}

\usepackage{tikz}
\usepackage{lipsum}
\makeatletter
\def\ieeecopyright{
    \footnotesize
    © 2025 IEEE. Personal use of this material is permitted.\newline
    DOI: 10.1109/RTSS62706.2024.00049}
\makeatother
\AddToHook{shipout/firstpage}{%
    \begin{tikzpicture}[remember picture,overlay]
        \node[anchor=south west,xshift=1.0cm,yshift=0.8cm]
        at (current page.south west)%
        {\parbox{\linewidth}{\raggedright\ieeecopyright}};
    \end{tikzpicture}%
}

\def\BibTeX{{\rm B\kern-.05em{\sc i\kern-.025em b}\kern-.08em
    T\kern-.1667em\lower.7ex\hbox{E}\kern-.125emX}}
\begin{document}

\title{Work-in-Progress: Multi-Deadline DAG Scheduling Model for Autonomous Driving Systems
    \thanks{This research is based on results obtained from Green Innovation Fund Projects / Development of In-vehicle Computing and Simulation Technology for Energy Saving in Electric Vehicles, JPNP21027, subsidized by the New Energy and Industrial Technology Development Organization (NEDO) and partially by JST PRESTO Grant Number JPMJPR21P1.}
}

\author{
    \IEEEauthorblockN{
        Atsushi Yano\IEEEauthorrefmark{1}\IEEEauthorrefmark{2},
        Takuya Azumi\IEEEauthorrefmark{1}\IEEEauthorrefmark{2},
    }
    \IEEEauthorblockA{\IEEEauthorrefmark{1}Graduate School of Science and Engineering, Saitama University, Japan}
    \IEEEauthorblockA{\IEEEauthorrefmark{2}TIER IV Incorporated, Japan}
}

\maketitle

\begin{abstract}
    Autoware is an autonomous driving system implemented on Robot Operation System (ROS) 2, where an end-to-end timing guarantee is crucial to ensure safety.
    However, existing ROS 2 cause-effect chain models for analyzing end-to-end latency struggle to accurately represent the complexities of Autoware, particularly regarding sync callbacks, queue consumption patterns, and feedback loops.
    To address these problems, we propose a new scheduling model that decomposes the end-to-end timing constraints of Autoware into local relative deadlines for each sub-DAG.
    This multi-deadline DAG scheduling model avoids the need for complex analysis of data flows through queues and loops, while ensuring that all callbacks receive data within correct intervals.
    Furthermore, we extend the Global Earliest Deadline First (GEDF) algorithm for the proposed model and evaluate its effectiveness using a synthetic workload derived from Autoware.
\end{abstract}

\vspace{-0.0mm}

\vspace{-0mm}

\section{Introduction}
\label{sec:introduction}
\vspace{-0mm}

Autonomous driving systems are garnering significant attention as potential solutions to social challenges such as mobility support in depopulated areas and logistics efficiency improvements, with research and development progressing globally.
In particular, Autoware~\cite{kato2018autoware} is an open-source autonomous driving system built on Robot Operation System (ROS) 2~\cite{teper2022end, teper2023timing, teper2024end}, and its modular architecture accelerates both the development and application of autonomous driving technologies.
Autoware performs essential autonomous driving functions, including sensing, localization, perception, planning, and control, using the publish/subscribe model with ROS 2 topics.
Nodes, which are the basic execution units in ROS 2, publish messages to topics, and other nodes subscribe to these messages for communication.
Each node uses ROS 2 callback mechanism to either wait for message arrivals or execute specific processes at fixed period.

Since autonomous driving systems are real-time systems where even slight delays can affect human lives, proper modeling and timing guarantees are required.
As the time from sensor data input to vehicle control command output is directly linked to the safety of autonomous driving systems, end-to-end timing guarantees are desirable.
To ensure end-to-end timing guarantees in complex systems, the cause-effect chain~\cite{gunzel2021timing, 10405874} is a model commonly used in real-time community.
Recent studies have proposed cause-effect chains that take into account the semantics of ROS 2, which can be used to derive the maximum reaction time and maximum data age for any end-to-end path within a ROS 2 system~\cite{teper2022end, teper2023timing, teper2024end}.

However, when applying the ROS 2 cause-effect chain to Autoware, three problems arise that cannot be adequately modeled, preventing accurate end-to-end timing guarantees: (i) the presence of sync callbacks, (ii) complex consumption mechanisms in message queues, and (iii) loop structures in the data flow.
While it is theoretically possible to reflect such problems in the cause-effect chain model, doing so would significantly increase the complexity of the model, complicating the analysis and requiring extensive research efforts to achieve timing guarantees.
As a result, it is currently challenging to establish design principles for schedulers aimed at improving the real-time performance of Autoware.

In this paper, we propose a new scheduling model for autonomous driving systems that balances ease of analysis and practicality, with a focus on short-term implementation.
The main idea is to decompose end-to-end timing constraints into local timing constraints for each partial Directed Acyclic Graph (DAG) and aim to guarantee the set of these local timing constraints.
This approach eliminates the need for analysis of complicated data flows involving queues and loops, while ensuring that all callbacks receive data within correct intervals.
Moreover, since our proposed model can be viewed as a specialization of the DAG tasks traditionally handled in real-time scheduling, it allows for the application of existing ideas from real-time DAG scheduling algorithms.

\textbf{Contributions:} The main contributions of this paper include the following:

\begin{itemize}
    \item We provide a detailed explanation of Autoware's callback graph and its complex trigger mechanisms, highlighting the discrepancies between ROS 2 cause-effect chains and Autoware.
    \item We propose a new multi-deadline DAG scheduling model that balances ease of analysis and empirical safety for autonomous driving systems.
    \item We extend the conventional Global Earliest Deadline First (GEDF) algorithm for DAG tasks to the proposed model and evaluate it using a synthetic workload that simulates Autoware.
\end{itemize}

\begin{figure*}[tb]
    \vspace{-3mm}
    \centering
    \begin{subfigure}{1.0\linewidth}
        \centering
        \includegraphics[width=1.0\linewidth]{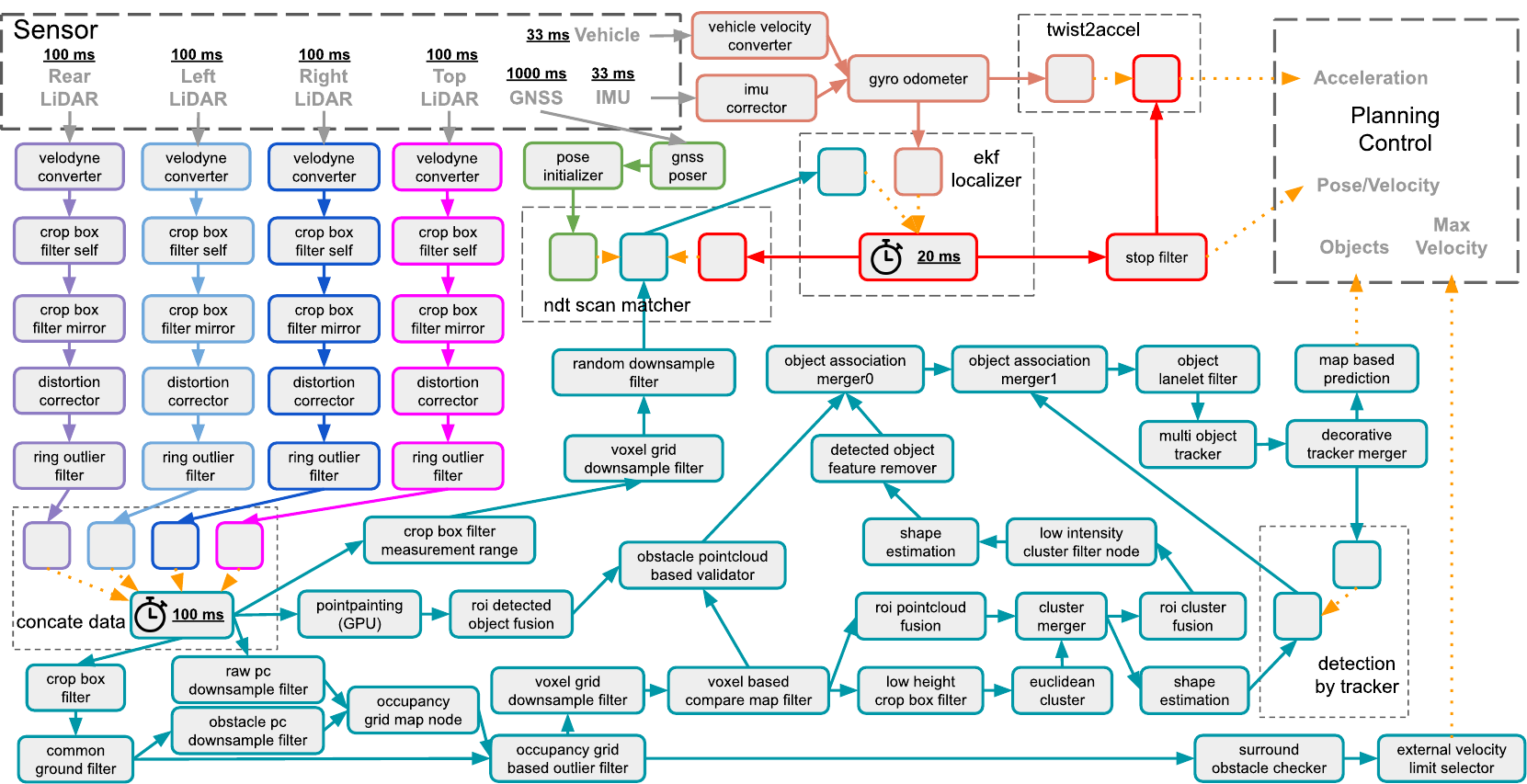}
    \end{subfigure}

    \vspace{0.3cm}

    \begin{subfigure}{0.85\linewidth}
        \centering
        \includegraphics[width=1.0\linewidth]{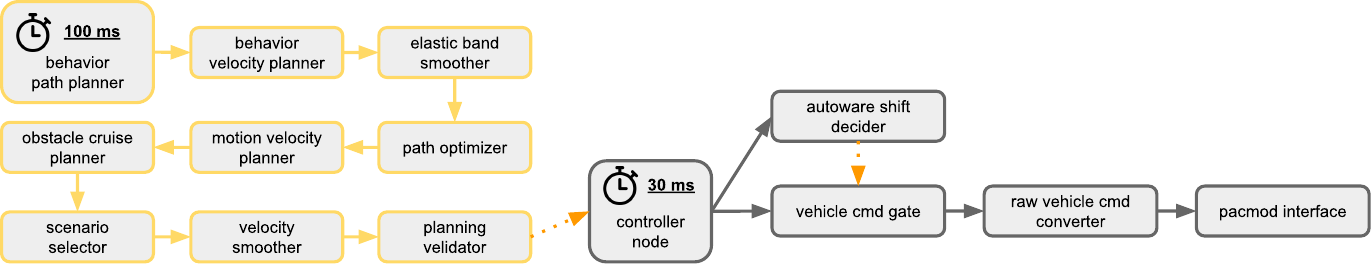}
    \end{subfigure}

    \caption{Main callbacks and their trigger mechanisms in Autoware.}
    \label{fig:autoware_dag}
    \vspace{-3mm}
\end{figure*}

\vspace{-0mm}
\section{Callback Graph and Trigger Mechanisms in Autoware} \label{sec:autoware}
\vspace{-0mm}

In this section, we describe the main ROS 2 callbacks that constitute Autoware and their triggering mechanisms, as illustrated in~\cref{fig:autoware_dag}.
Autoware is a system that performs sensor data processing, self-localization, object detection, route design, and vehicular behavior decisions, using multiple sensors operating at unique periods (e.g., LiDAR, GNSS, and IMU) as data sources, and ultimately outputting appropriate vehicle control commands.
The gray rectangles in~\cref{fig:autoware_dag} represent individual ROS 2 callbacks, and when there is only one callback within a ROS 2 node, it represents the node itself.
On the other hand, if a node contains multiple callbacks, a gray dotted square surrounds the set of callbacks, indicating the node (e.g., \textit{ndt\_scan\_matcher} and \textit{concat\_data}).

The solid edges indicate publish/subscribe communication via the same topic, while the orange dotted edges represent data transfer through the queue in node member variables or via the ROS 2 take API~\cite{takeAPI}.
The take API allows node to receive messages from a queue of the subscribed topic at any time without using a callback.
Therefore, the orange dotted edges do not directly trigger subsequent callbacks but influence the processing time and output accuracy of those callbacks.
For instance, the \textit{ndt\_scan\_matcher} receives the previous self-localization result from the \textit{ekf\_localizer}, stores it as a member variable, and uses it as the initial pose for the next scan matching~\cite{ndt_scan_matcher}.

The callbacks used in Autoware can be classified into the following three categories based on their trigger conditions:

\begin{itemize}
    \item \textbf{Timer Callback}: A callback triggered at a fixed period, represented by a timer icon in~\cref{fig:autoware_dag}.
    \item \textbf{Subscription Callback}: A callback triggered by a publish for a single topic that is subscribed to.
    \item \textbf{Sync Callback}: A callback triggered by the synchronization of data from multiple subscribed topics using the ApproximateTime policy~\cite{9984711} in ROS 2 (e.g., \textit{occupancy\_grid\_map\_node} and \textit{cluster\_merger}).
\end{itemize}

\noindent The reason for using timer callbacks in Autoware is to ensure that data is output at periods required by subsequent callbacks, regardless of when the data is received.
For example, the \textit{ekf\_localizer} publishes self-localization data at 20 ms periods to subsequent callbacks by using a Kalman filter~\cite{10.1115/1.3662552} to interpolate the outputs of the \textit{ndt\_scan\_matcher}, which arrive approximately every 100 ms.

\section{Problem Statement} \label{sec:problem_statement}

In this section, we discuss the challenges of modeling Autoware using cause-effect chains.
While end-to-end timing guarantees are a promising option, state-of-the-art ROS 2 cause-effect chains~\cite{teper2022end,teper2023timing,teper2024end} cannot properly model Autoware due to the following problems.

\textbf{[Problem 1] Cause-effect chains cannot model sync callbacks: }
\noindent Even the latest literature on ROS 2 cause-effect chains assumes the existence of only one publisher per subscription, but Autoware contains many sync callbacks.

\textbf{[Problem 2] Data in queues is not always consumed in a FIFO manner: }
\noindent In communication between callbacks via queues, it is often assumed for simplification that the oldest data is consumed exactly once.
However, this is not always the case in Autoware.
For example, there are callbacks that consume all the data present in the queue at that moment, while others consume data selected based on timestamps rather than the oldest data.

\textbf{[Problem 3] Handling feedback from descendant callbacks is non-trivial: }
\noindent Autoware has callbacks that generate output by referencing feedback from descendant callbacks (e.g., \textit{ekf\_localizer} to \textit{ndt\_scan\_matcher}).
How to handle such cases in cause-effect chains is not straightforward.

Considering the problems above, the complexity of the cause-effect chain model and the difficulty of end-to-end timing analysis increase significantly.
Furthermore, developing a scheduler that minimizes the maximum reaction time and the maximum data age based on this end-to-end timing analysis will require further research efforts, and its practical implementation will take considerable time.

\section{Proposed Model and Extended EDF} \label{sec:proposal}

To address the problems presented in Section~\ref{sec:problem_statement}, we propose an approach that models Autoware as DAG tasks by fully segmenting the graph at the orange dotted edges shown in~\cref{fig:autoware_dag}.
This approach aims to eliminate the need for analyzing complex communications via queues by decomposing end-to-end timing constraints into local timing constraints, i.e., the relative deadline for each sub-DAG.
In other words, if the relative deadlines of all DAGs are satisfied, the entire system is considered safe.
This assumption is based on the empirical knowledge that as long as each callback receives data from all predecessor callbacks within a predetermined interval (relative deadline for prior DAGs), the system operates safely.

By splitting Autoware at the orange dotted lines, callbacks of the same frame color in~\cref{fig:autoware_dag} are considered to belong to the same DAG.
As depicted in ~\cref{fig:autoware_dag}, a single DAG in the proposed model has one more sink callback, which send data to different subsequent callbacks.
According to our assumption, since the receiving callbacks are different, their data request intervals vary, meaning that each sink callback has its own distinct relative deadline.
Here, the relative deadlines for each DAG must be set by the system designer; however, the method for determining these values is beyond the scope of this paper.

We now explain how the proposed model addresses the problems presented in Section~\ref{sec:problem_statement}.
For Problem 1, a sync callback in a typical configuration can be considered as a precedence constraint in the traditional DAG models used in real-time scheduling~\cite{verucchi2023survey}.
For Problems 2 and 3, the proposed model abstracts communications via queues as the relative deadline of each DAG.
In other words, proposed model only guarantees that each DAG outputs data within a specified interval, and how the data propagates afterward is outside the scope of the analysis.

\subsection{Formalization of Proposed Model}

This section formalizes of the proposed model, which involves splitting Autoware into the DAG tasks.
The DAG task set is denoted by $\tau \triangleq \{\tau_1, \ldots, \tau_n\}$, where $n$ is the total number of tasks.
Each task is defined by three-tuple: $\tau_x \triangleq (T_x, G_x, \mathfrak{D}_x)$.
$T_x$ represents the period of $\tau_x$.
The DAG corresponding to task $\tau_x$ is represented as $G_x \triangleq (V_x, E_x)$, where $V_x$ is the set of vertices and $E_x$ is the set of edges in the DAG.
Each vertex in $V_x$, denoted as $v_{x,i}$, while the edges in $E_x$ represent the precedence constraints between these vertices.
Additionally, $des(v_{x,j})$ refers to the set of descendant vertices of $v_{x,j}$.

We assume that each DAG has a single source vertex and one or more sink vertex with a unique relative deadline.
The set of sink vertices in $G_x$ is denoted by $sink_x$.
The set $\mathfrak{D}_x$ contains the relative deadlines for $G_x$, where each relative deadline specifies a timing constraint from a source vertex to a sink vertex.
Specifically, $D_{x,i} \in \mathfrak{D}_x$ represents the relative deadline from the source vertex to $v_{x,i} \in sink_x$.

Each task $\tau_x$ continually generates the job instances, and $J^k_x$ represents the $k$-th job instance of task $\tau_x$.
From this point on, symbols with a superscript $k$ indicate that they are associated with $k$-th job instance.
The set $\mathfrak{d}^k_x = \{D_{x,i} + kT_x \mid D_{x,i} \in \mathfrak{D}_x\}$ represents the absolute deadlines for $J^k_x$.
The absolute deadline from the source vertex to the sink vertex \(v^k_{x,i}\) in $J^k_x$ is denoted by \(d^k_{x,i} \in \mathfrak{d}^k_x\).

\subsection{Extended EDF Algorithm} \label{subsec:extended_edf}

In this section, we provide a simple extension of Global Earliest Deadline First (GEDF) for the proposed model.
In conventional GEDF for DAG tasks, all nodes within a single DAG task job share the same priority, as they all share the job's absolute deadline~\cite{6602083}.
However, in the proposed model, it is necessary to account for the fact that multiple deadlines can exist within a single DAG.
Therefore, we propose to use the Reference Absolute Deadline (RAD) as the priorities in GEDF, defined as follows:

\vspace{2mm}
\begin{math}
    \textrm{RAD}(v^k_{x,i}) \triangleq
\end{math}
\begin{align*}
    \begin{cases}
        d^k_{x,i}                                                     & \textrm{if} \quad v_{x,i} \in sink_x \\
        \min(\{d^k_{x,j} \mid v_{x,j} \in des(v_{x,i}) \cap sink_x\}) & \textrm{otherwise}.
    \end{cases}
\end{align*}

\noindent This extension indicates to use the earliest time among the absolute deadlines of all descendant sink vertices when a given vertex is not a sink vertex.
In this way, by using the proposed model, existing ideas from real-time DAG scheduling algorithms can be leveraged, making it easier to design schedulers aimed at minimizing deadline misses.

\section{Evaluation for Extended GEDF} \label{sec:evaluation}

In this section, we conduct a fundamental experiment of the extended GEDF algorithm proposed in Section~\ref{subsec:extended_edf} via a scheduling simulation using a synthetic workload based on Autoware.
The following two algorithms are used for comparison: (i) work-conserving scheduler~\cite{melani2015response} and (ii) Rate Monotonic (this is also work-conserving) scheduler (RM).

In the workload generation process, the dependencies and periods of the callbacks are set as shown in~\cref{fig:autoware_dag}, and these are kept constant throughout the experiment.
The relative deadline for each sink vertex is determined based on (worst-case execution time-based) critical path length to its sink vertex, which is also fixed during the experiment.
The execution time for each callback is randomly selected from the actual measurements obtained by CARET~\cite{CARET}, and during the simulation, all callbacks are executed for the fixed time that is selected.
The same DAG tasks is used for all algorithms to ensure consistency.

The simulator implements a homogeneous multi-core processor, and since the total utilization (i.e., the sum of the total execution time divided by the period of each DAG) during the experiments never exceeded seven, the number of cores is fixed at seven.
The simulation duration is set to 3,000 ms, which corresponds to the hyper-period of Autoware DAG tasks.
For each algorithm, 5,000 simulations are conducted.

The evaluation metric used is the acceptance ratio, which is defined as the number of simulations completed without deadline misses divided by the total number of simulations.
The normalized utilization is calculated by dividing the total utilization of the DAG tasks used in the simulations by the number of cores (i.e., seven).
This value is then rounded to the nearest 0.5 units, and the acceptance ratio is computed for each point.
The program used for the evaluation is publicly available\footnote{\url{https://github.com/atsushi421/2024_RTSS_WiP_Evaluation}}.

\begin{figure}[tb]
    \centering
    \includegraphics[width=1.0\columnwidth]{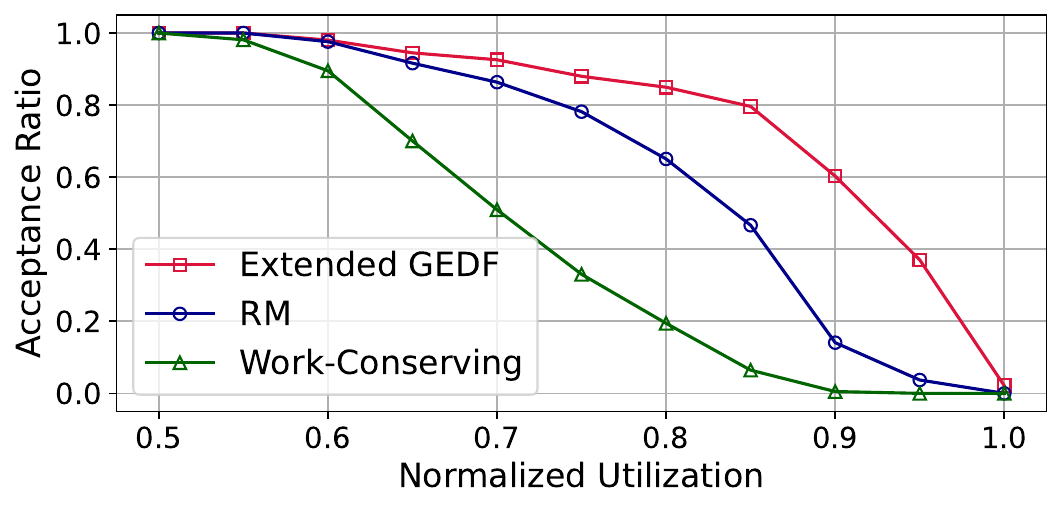}
    \caption{Acceptance ratio result of the extended GEDF and the two baseline algorithms.}
    \label{fig:acceptance_ratio}
\end{figure}

From the results shown in~\cref{fig:acceptance_ratio}, the extended GEDF outperforms all other algorithms in every case.
This is due to the nature of EDF, which minimizes deadline misses by prioritizing the vertex with the earliest deadline, and the extended EDF demonstrates that it is able to inherit this nature appropriately.

\section{Conclusion} \label{sec:conclusion}

This paper has proposed a multi-deadline DAG scheduling model tailored for Autoware addressing the limitations of existing cause-effect chain models.
By decomposing end-to-end timing constraints into local relative deadlines for each sub-DAG, the proposed model simplifies analysis, eliminating the need to account for data flow complexities caused by queues and loops.
We have also extended the GEDF algorithm to fit this model, demonstrating its effectiveness through simulations and how seamlessly the proposed model integrates with existing real-time DAG scheduling algorithms.
As future work, we plan to investigate methods for appropriately determining the relative deadline for each DAG and analyze how guaranteeing local timing constraints affects the measured end-to-end reaction times and data age.


\bibliographystyle{bibtex/IEEEtran}
\bibliography{bibtex/master}

\end{document}